# Interference and Efficient Transmission Range via V2V Communication at Roads Traffic Intersections

Ala Alobeidyeen, and Lili Du (lilidu@ufl.edu, Corresponding Author)

Civil and Coastal Engineering Department, University of Florida

*Abstract*— Vehicle-to-Vehicle (V2V) communication technology has dramatically promoted many promising applications to enhance traffic safety, mobility, and sustainability. However, However, we still lack the understanding of some fundamental properties of V2V technology under urban traffic conditions, such as interference at traffic intersections. Motivated by this view, this study develops the mathematical formulations to capture the worst-case interference at traffic intersections, considering the macroscopic traffic flow conditions and critical road geometric features including intersection diameter D, and intersection angle α. Built upon these formulations, we develop a mathematical model to approximate a conservative transmission range to sustain the successful V2V transmission at a traffic intersection. Our experiments illustrate that the proposed analytical formulations can provide accurate approximations for the interference and the corresponding transmission range at orthogonal (non-orthogonal) traffic intersections under various traffic congestion levels. Furthermore, this study conducted other experiments to understand how intersection geometric features (such as (D, α)) impact V2V communication at traffic intersections. The results illustrate that severer interference and smaller transmission range occur at a smaller intersection (with smaller diameter D) under heavy traffic congestion level. And the orthogonal intersection (i.e. with $α=90^0$) gives critical thresholds (such as severest interference and minimum transmission range) under all different traffic conditions, which help in understanding the V2V communication performance at an urban traffic intersection. These findings will potentially help to develop efficient MAC algorithms adaptive to urban traffic conditions, and further support various ITS applications using V2V communication. interference and transmission range.

## I. INTRODUCTION

Vehicle-to-Vehicle (V2V) communication network, also named as VANET, mainly operates in 5.85-5.92 GHz DSRC spectrum, which is allocated by the Federal Communications Commission (FCC) to support the Intelligent Transport System (ITS) applications in the short-range communication in vehicular networks [1]. To promote the applications of the V2V communication based upon DSRC, the IEEE 802.11p communication protocol is approved as an amendment of the Wi-Fi communications standards to serve V2V/V2I communications in the 5.9 GHz DSRC spectrum under a vehicular environment [2], in which the MAC layer is designed to regulate the access of the V2V transmissions to the shared communication channel in the PHY layer so that the interference and collisions can be minimized, while the PHY is more closely associated with traffic stream features.

Many national and international projects, such as CarTALK2000 [3], Car2Car Communication Consortium (C2CCC) [4], California Partners for Advanced Transit and Highways (PATH) [5], and FleetNet [6], tested the applicability of V2V technologies in various transportation scenarios. It has been well accepted that VANET has a great potential to improve traffic safety [7][8], mobility [9], and sustainability [10], but still faces many implementation challenges, which hold the application of these promising innovations back from success. For example, it has been noticed that the signal interference issue caused by dense traffic at intersections or on highways may significantly affect V2V communications [11][12][13]. It represents the main factor to cause package drop, delay or other types of communication failures [14][15].

Signal interference is one of the critical issues of V2V communication under urban traffic environment, many research efforts in literature have investigated these issues from different perspectives. Scholars in wireless communication community seek to address it by developing various adaptive MAC schemes, such as [16]–[21]. However, the traffic congestion especially that at intersections has been only factored by a few parameters such as traffic density; dynamic traffic flow evolution has never been considered. On the other hand, researchers in transportation community demonstrate more interests in understanding the connectivity of VANET considering the properties of the PHY layer associated with traffic flow dynamics, such as [22]–[24]. Many analytical models have been developed to capture information dissemination via V2V on a road segment, such as [15][25][26][27], without factoring the impact of the cumulative interference on the efficient transmission range. A few recent studies in [22][28] tackled the interference problem on roadways, but their formulations are not adaptive to road intersections.

The state of the art demonstrates that neither the signal interference via V2V at traffic intersections, nor the corresponding efficient transmission range has been well explored by mathematical analysis. Wireless communication and transportation communities either ignore or oversimplify traffic features at intersections, where severe interference often cause transmission failure [29] due to dense traffic. Moreover, the well-known Signal-to-Interference Noise Ratio (SINR), developed by Gupta and Kumar in [30], is employed to judge the success of a wireless communication, factoring the signal interference in VANET. The employed theoretical formulation for calculating the interference requires the spacing measurements between the receiver and each other transmitter (a vehicle). This study notices the weakness of this interference formulation as follows. First, these data vary quickly under dynamic urban traffic conditions and they are often difficult to obtain. In addition, the spacing between a receiver and the other vehicles cannot reflect the traffic condition well, such as congestion levels. Therefore, the interference formulation used in the SINR is not well adaptive to understand the signal interference under urban traffic conditions. Hence, it cannot facilitate the design of MAC protocols to fit well under possible traffic condition.

Motivated by the above view, our study is dedicated to bridge these methodological gaps by developing mathematical models to capture the interference at traffic intersection factoring multiple macroscopic traffic characteristics including average traffic spacing headway and critical intersection geometric features. Built upon these models, we further develop a conservative upper bound of transmission range incorporating the signal interference associated with dynamic traffic features at intersection. The outcomes of this study will potentially sustain the usage of V2V communication under urban environments in various ITS applications.

The efforts of this study are presented by the structure as follows. Section I introduces research background, motivations and relevant research in literature. Section II presents the problem formulations, including the necessary background of V2V successful wireless communication condition; Section III proposes our methodology to develop mathematical formulations to capture the signal interference and the conservative transmission range at traffic intersection. The proposed formulations are validated by numerical experiment tests presented in Section IV, and the conclusion of this study is given in Section V.

## II. PRELIMINARIES

This research considers a fleet of vehicles that are equipped with 5.9 GHz DSRC wireless communication facilities so that they can conduct V2V communication. The standard SINR model, given in Eq. (1) is used to identify a successful communication between a transmitter (such as vehicle w) and a receiver (vehicle i) in the physical layer, factoring signal interference [30]. It indicates that a successful wireless communication occurs if the value of SINR is greater than a threshold β.

$$\text{SINR} = \frac{P_w (x_{wi})^{-\gamma}}{N + I} \geq \beta; \quad I = \sum_{j=1, j \neq w}^{n} e_j P_j (x_{ji})^{-\gamma}, \quad (1)$$

where $P_w$ represents the transmission power of node $w$; $x_{wi}$ represents the distance between transmitter vehicle $w$ and receiver vehicle $i$; $\gamma$ is the signal power decay, typically $2 \leq \gamma \leq 6$; $N$ represents the background noise on the frequent channel utilized by network; $\beta$ is the threshold which depends on the designing modulation and code rate (value which indicates the data transmission rate during a wireless connection) [31]; $I$ represent the sum of interference power reaching to vehicle $i$ from all other vehicles ($j \neq w$) except the transmitter $w$. $e_j = 1$, if vehicle j is in transmission status; otherwise, $e_j = 0$.

The interference formulation $I$ employed in Eq. (1) indicates that three main quantitative measurements will affect the signal interference under urban transportation environments. They are the transmission power (i.e., parameter $P_w$), MAC algorithm (i.e., parameter $e_j, \forall j$), and the distribution of neighborhood vehicles around the receiver (i.e., spacing $x_{ji}, \forall j \neq w$). To investigate the signal interference at traffic intersection, we make the following assumptions for the transmission power and MAC protocol. We consider all vehicles adopt the same transmission power (i.e., $P_w = P_j = P, \forall j$) and broadcast messages by flooding (i.e., $e_j = 1, \forall j$). Thus, the interference measured by this study represents the worst case.

Moreover, this study notices that the spacing information used in formulation $I$ refers to microscopic vehicle trajectory data, which are expensive and vary quickly due to high traffic dynamics. Moreover, the geometric features of the crossing roadways at traffic intersection will also affect the measurement of signal interference. These features are not well involved in the interference formulation $I$ for measuring the SINR. These weaknesses make the SINR is not sufficient to understand the performance of the V2V in urban traffic intersections under dynamic traffic conditions. The above view motivates this study to explore a new interference formulation, which is adaptive to urban traffic intersections better.

## III. METHODOLOGY

To The study seeks to develop a new interference formulation integrating macroscopic traffic flow characteristics (such as average headway spacing h, and geometric features (D, α)[1]) at an intersection (see Fig. 1). These characteristics unlike the microscopic trajectory data, are not only conventionally identified traffic congestion conditions (such as sparse, mild, and heavy congestion level), but also relatively easy to be collected and tracked. The methodology is developed by taking an urban orthogonal intersection with four arms as an example, and further extended to more general traffic intersections. The formulations developed in this study seek to help understand the average interference under different traffic congestion scenarios, which are significantly important for the design of MAC algorithms in order to fit well under possible traffic conditions. On top of that, it will help identify the success of V2V transmissions at traffic intersections.

---
[1] **D** is the intersection diameter between two furthest opposing arms corners, and **α** is the intersection angle between two adjacent arms.

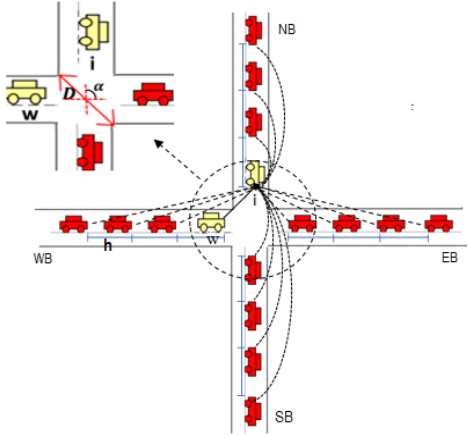

Fig. 1. Illustration for Transmission Interference at vehicle i (Intersection mode).

*A. Interference upper bound at traffic intersection:*

According to the assumptions for the transmission power and MAC, the interference formulation $I$ in the SINR can be transferred to Eq. (2), which selects the decay rate exponent $\gamma=2$, corresponding to free space[2] information propagation [32]; and take off the constant parameter associated to the transmission power. Eq. (2) provides the interference surrogate occurring at the receiver vehicle i. Built upon that, the study below will develop more comprehensive formulations to capture the interference at traffic intersections.

$$\Lambda_i = \sum_{j=1, j \neq w}^{n} (x_{ji})^{-2}, \quad (2)$$

First of all, this study estimates the spacing (i.e., $x_{ji}$) for each individual vehicle j located on intersection arms to the receiver I by using both headway h and geometric features $(D, \alpha)$ at an intersection. Taking a 4-arm orthogonal intersection with a single lane in **Error! Reference source not found.** as an example, we demonstrate our formulations in Eqs. (3) and (4). For example, the distance $(x_{ji})$ for each j vehicle on the west and east arms in Fig. 2(a) is computed using a trigonometric function (i.e. third side of a triangle and law of cosine) which involves the spacing h between each individual vehicle, the intersection diameter D, and the angle α. Similarly, we compute distance $(x_{ji})$ for j vehicles on the south and north arms using the spacing h and the diameter D, as shown in Fig. 2(b).

Moreover, we approximate each spacing h by the average headway value, which presents the macroscopic traffic conditions and also simplifies the formulations.

$$(x_{ji})^2_{EB/WB} = \frac{D^2}{4} + (\frac{D}{2} + jh)^2 - 2(\frac{D}{2})(\frac{D}{2} + jh)\cos\alpha \quad (3)$$

$$(x_{ji})^2_{SB/NB} = D^2 + (D + jh)^2 \quad (4)$$

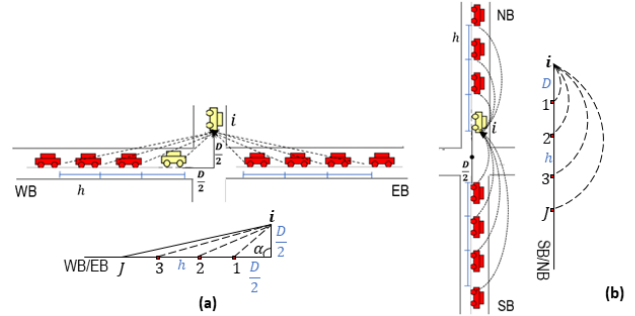

Fig. 2. Illustration for (a) distance $(x_{ji})$ from j vehicles on WB/EB to receiver i (b) distance $(x_{ji})$ from j vehicles on SB/NB to receiver i.

Given the receiver i locating at the stop line of the northbound, the interference $\Lambda_i$ shown in Eq. (5) involves the effects from vehicles located at the north, south, east, and west arms.

$$\Lambda_i(x) = (\Lambda_i)_N + (\Lambda_i)_S + (\Lambda_i)_E + (\Lambda_i)_W. \quad (5)$$

By using trigonometric functions and approximating the spacing between adjacent vehicles on each arm by the average headway h (i.e., $(x_{ji})$), we develop the closed-formulations for $(\Lambda_i)_N, (\Lambda_i)_S, (\Lambda_i)_E$, and $(\Lambda_i)_W$ in Eqs. (6), (7), (8), and (9) involving the geometric features at the traffic intersection.

$$(\Lambda_i)_N = \sum_{j=1}^{n_N} 1/(x_{ji})^2 = [\frac{1}{h^2}\sum_{j=1}^{n-1}\frac{1}{j^2}], \quad (6)$$

$$(\Lambda_i)_S = \sum_{j=1}^{n_S} 1/(x_{ji})^2 = [\frac{1}{D^2} + \sum_{j=1}^{n_S-1}\frac{1}{(D+jh)^2}], \quad (7)$$

$$(\Lambda_i)_W = \sum_{j=1}^{n_W} 1/(x_{ji})^2 = [\frac{1}{\frac{D^2}{4}(2-2\cos\alpha)}\sum_{j=1}^{n_W-1}\frac{1}{(jh)^2+(jh)(D)-(jh)(D)\cos\alpha}], \quad (8)$$

$$(\Lambda_i)_E = \sum_{j=1}^{n_E} 1/(x_{ji})^2 = \frac{1}{\frac{D^2}{4}(2-2\cos\alpha)} + [\frac{1}{\frac{D^2}{4}(2-2\cos\alpha)}\sum_{j=1}^{n_E-1}\frac{1}{(jh)^2+(jh)(D)-(jh)(D)\cos\alpha}], \quad (9)$$

where $n_*$ represents the number of vehicles involved in each arm; $(\Lambda_i)_*$ represents the interference coming from each arm. Clearly, $(\Lambda_i)_*$ is affected by the geometric features and the traffic distribution around the traffic intersection. Through processing Eqs. (6)–(9), we have the following Proposition:

**Proposition 1:** Assuming that the arms segment with very long length, so that the number of vehicles $n_* \to \infty$, we have:

---

[2] **γ** typically varies between 2 to 6 and equal 2 under free space. We choose γ=2 considering it is suited to both open highway and urban intersection scenarios. Note that we consider that there is no obstruction between a receiver and transmitter located at stop line.

$$\Lambda_i(x) = (\Lambda_i)_N + (\Lambda_i)_S + (\Lambda_i)_E + (\Lambda_i)_W \leq \frac{\pi^2}{6h^2} + \left[\frac{1}{D^2} + \frac{1}{\frac{D^2}{2}(1-\cos\alpha)}\right] + \left[\frac{1}{h^2} \times \Psi_1\left(\frac{D}{h}\right)\right] + \frac{1}{\frac{1}{2}(D)^3 h(1-\cos\alpha)^2} \times \left(\Psi\left(\frac{D}{h}(1-\cos\alpha)\right) + \frac{1}{\frac{D}{h}(1-\cos\alpha)} + \gamma\right)^2. \quad (10)$$

Where,
a. $\gamma$ is the Euler-Mascheroni constant approximate equals 0.5772,
b. $\Psi(z)$ and $\Psi_0(z)$ are respective digamma function and the logarithmic derivative of the gamma function. Namely, $\Psi_0(z) = \frac{\Gamma'(z)}{\Gamma(z)}$.
c. $\Psi_n(z+1)$ is polygamma function [z+1, n]. The $n^{th}$ logarithmic derivative of the gamma function, where (n) should be non-negative integer order. Also, in this function the notation of $\Psi_0(z) = \Psi(z)$,

**Proof:** First, according to [33], we know $\sum_{j=1}^{\infty}\frac{1}{j^2} = \frac{\pi^2}{6}$. Thus, $(\Lambda_i)_N$ in Eq. (6) has an upper bound shown in Eq. (11).

$$(\Lambda_i)_N \leq \left[\frac{1}{h^2} \times \sum_{j=1}^{\infty}\frac{1}{j^2}\right] \leq \frac{\pi^2}{6h^2} \quad (11)$$

We next take a look at $(\Lambda_i)_S$ in Eq. (7). According to [34], we have $\sum_{j=1}^{\infty}\frac{1}{(D+jh)^2} = \frac{1}{h^2}\sum_{j=1}^{\infty}\frac{1}{(\frac{D}{h}+j)^2} = \frac{1}{h^2}\zeta\left(2,\frac{D}{h}\right)$, where $\zeta\left(2,\frac{D}{h}\right)$ is the Hurwitz zeta function. Furthermore, according to [35], we have $\zeta\left(2,\frac{D}{h}\right) = \Psi_1\left(\frac{D}{h}\right)$, where $\Psi_n(z)$ is the polygamma function[3]. Under these approximations, we obtain the upper bound of $(\Lambda_i)_S$ in Eq. (12).

$$(\Lambda_i)_S \leq \frac{1}{D^2} + \frac{1}{h^2} \times \Psi_1\left(\frac{D}{h}\right) \quad (12)$$

$(\Lambda_i)_W$ in Eq. (8) can also be processed using the similar approach. It is noticed that $\sum_{j=1}^{\infty}\frac{1}{(jh)^2+(jh)(D)-(jh)(D)\cos\alpha} \times \frac{\left(\frac{1}{h}\right)^2}{\left(\frac{1}{h}\right)^2} = \frac{1}{h^2} \times \sum_{j=1}^{\infty}\frac{1}{j\left[j+\frac{D(1-\cos\alpha)}{h}\right]}$. According to [36], we have $\sum_{j=1}^{\infty}\frac{1}{j\left[j+\frac{D(1-\cos\alpha)}{h}\right]} = \frac{\Psi\left(1+\frac{D(1-\cos\alpha)}{h}\right)}{\frac{D}{h}(1-\cos\alpha)} + \frac{\gamma}{\frac{D}{h}(1-\cos\alpha)}$, where $\Psi(z+1)$ is digamma function [4]. Also, according to [36], $\Psi(1+z) = \Psi(z) + \frac{1}{z}$. Under these approximations, we obtain the upper bound of $(\Lambda_i)_W$ in Eq. (13).

$$(\Lambda_i)_W \leq \frac{1}{\frac{1}{2}(D)^3 h(1-\cos\alpha)^2} \times \left[\Psi\left(\frac{D}{h}(1-\cos\alpha)\right) + \frac{1}{\frac{D}{h}(1-\cos\alpha)} + \gamma\right]. \quad (13)$$

Finally, it is noticed that $(\Lambda_i)_E$ in Eq. (9) can be re-written as $(\Lambda_i)_E = \frac{1}{\frac{D^2}{4}(2-2\cos\alpha)} + (\Lambda_i)_W$. Thus, we obtain the upper bound of $(\Lambda_i)_E$ in Eq. (14) using the results for $(\Lambda_i)_W$ in Eq. (13).

$$(\Lambda_i)_E \leq \frac{1}{\frac{D^2}{4}(2-2\cos\alpha)} + \frac{1}{\frac{1}{2}(D)^3 h(1-\cos\alpha)^2} \times \left[\Psi\left(\frac{D}{h}(1-\cos\alpha)\right) + \frac{1}{\frac{D}{h}(1-\cos\alpha)} + \gamma\right] \quad (14)$$

Summing Eqs. (11)–(14), this study obtains the upper bound of $\Lambda_i$ given in Eq. (10), which only involves the macroscopic traffic features (spacing headway h) and geometric features of the 4-arms orthogonal intersections. This closes the proof of the Proposition.

*B. Interference upper bound for orthogonal intersection:*

The upper bound of the interference surrogate given in Eq. (10) does not present the merits in applications since the polygamma and digamma functions, i.e., $\Psi\left(\frac{D}{h}(1-\cos\alpha)\right)$ and $\Psi_1\left(\frac{D}{h}\right)$, do not have closed-form expressions. This study thus considers developing an approximation for the upper bound. Taking an orthogonal intersection ($\alpha = 90°$) with four single-lane arms as an example, we demonstrate our approach as follow. First, given $\alpha = 90°$, we can simplify the upper bound of the interference surrogate shown in Eq. (10) to Eq. (15).

$$\Lambda_i(x) \leq \frac{3}{D^2} + \frac{1}{h^2}\left[\frac{\pi^2}{6} + \Psi_1\left(\frac{D}{h}\right)\right] + \left[\frac{2}{D^3 h} \times \left(0.5772 + \frac{h}{D} + \Psi\left(\frac{D}{h}\right)\right)^2\right] \quad (15)$$

Next, this study notices that the polygamma and digamma functions in Eq. (10) depend on the geometric features, such as the value of D and $\alpha$, and the traffic distribution such as the spacing (h) at an intersection. The feasible values of these parameters are limited by the roadway design and traffic reality. Previous research in [37] and [38] observed that the minimum safe spacing (h) for vehicles approaching an urban intersection equal 86 ft. And, the traffic flow data [39] shows that the smallest spacing headway (h) could approximately equal 1.5 ft in very congested traffic condition. Thus, this study considers h∈ [1.5, 86] ft, representing a reasonable range for the spacing between vehicles at traffic intersections. Moreover, the study in [38] demonstrates that the street width per lane is from 20 ft to 88 ft in the six to eight-lanes undivided streets. It indicates that D∈ [28, 125] ft. represents a possible range for the parameter. Accordingly, this study finds that $\left(\frac{D}{h}\right) \in [0.33, 83.33]$ holds.

With the knowledge of the geometric features of traffic intersections, we investigate the curves of the polygamma and digamma functions within the range of the $\left(\frac{D}{h}\right)$ ratio. The regression analysis shown in Fig. 3 and Fig. 4 demonstrate that the power and logarithmic distributions respectively, fit the polygamma and digamma data very well. Consequently, substituting the polygamma and digamma functions by their approximation functions in Eq. (15), we obtain Eq. (16) to estimate the upper bound of $\Lambda_i$.

---

[3] $\Psi_n(z) = (-1)^{n+1} \times n! \, \zeta(n+1, z)$

[4] $\Psi(1+z) = -\gamma + \sum_{n=1}^{\infty}\frac{z}{n[n+z]}$

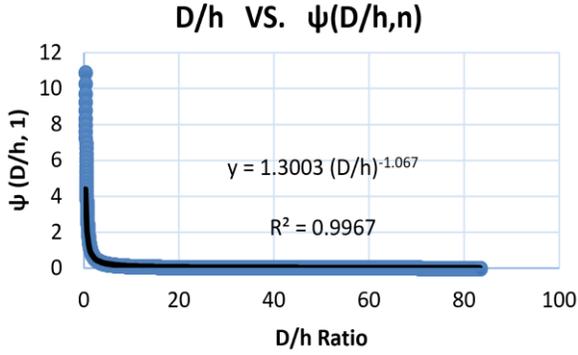

Fig. 3. $\left(\frac{D}{h}\right)$ ratio vs. Polygamma function ψ(D/h,1) power-distribution fits the data, by producing exponents -1.067

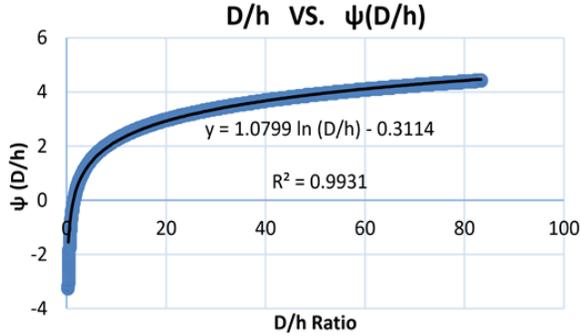

Fig. 4. $\left(\frac{D}{h}\right)$ ratio vs. Digamma function ψ(D/h): logarithmic distribution fits the data.

$$\Lambda_i(x) \leq \frac{3}{D^2} + \frac{1}{h^2}\left[\frac{\pi^2}{6} + 1.3003\left(\frac{D}{h}\right)^{-1.067}\right] + \left[\frac{2}{D^3 h} \times \left(0.2658 + \frac{h}{D} + 1.0799 \ln\left(\frac{D}{h}\right)\right)^2\right] \quad (16)$$

Note that the formulation in Eq. (16) only provides the upper bound of the interference surrogate $\Lambda_i$. Under different congestion conditions and traffic intersections (i.e., different values of h and D), the corresponding upper bounds vary. Given the feasible values of h and D are limited by traffic flow reality and intersection design codes, this study thus thinks about exploring the tightest upper bound of $\Lambda_i$. To do that, we form a mathematical programming (MP) in Eq. (17), which seeks to minimize the interference upper bound in Eq. (16) subject to realistic values of D and h. The MP is a nonlinear program with simple box constraints for two variables. Thus, it can be easily solved by performing grid search.

$$\Lambda_{min} = \text{Min } \Lambda_i(x) = \frac{3}{D^2} + \frac{1}{h^2}\left[\frac{\pi^2}{6} + 1.3003\left(\frac{D}{h}\right)^{-1.067}\right] + \left[\frac{2}{D^3 h} \times \left(0.2658 + \frac{h}{D} + 1.0799 \ln\left(\frac{D}{h}\right)\right)^2\right] \quad (17)$$

s.t:
$1.5 < h \leq 86,$
$28 \leq D \leq 125,$

### C. Interference upper bound for non-orthogonal intersection:

This study next addresses the case of a general traffic intersection. Similar to the orthogonal intersection, we develop an approximation for the upper bound of the interference surrogate given in Eq. (10) for a non-orthogonal intersection with angle α ≠ 90°. To do that, first we noticed that the macroscopic traffic parameter of the average spacing headway (h), and the range for the parameter (D) are still applicable at the traffic intersection. However, to address various potential cos(α) of the interference surrogate formulation in Eq. (10), this study considers the road intersection design code to represent a reasonable range for angle (α) at non-orthogonal intersections. More exactly, according to AASHTO policy [40], the intersection angle (α) varies within the range of [60°, 90°]. Moreover, only the values providing the minimum adequate sight distance at safe intersection are applicable in practice. Following these engineering codes, we demonstrate our approach to enumerate the realistic values of α using the case in Fig. 5. as an example. The sight distance at a traffic intersection is restricted by three-triangle legs (a, b, c) which are directly related to vehicle speed limit and braking distance traversed by vehicle during perception and reaction time [40]. The two legs (a, b) can be calculated using the formulation in Eq. (18).

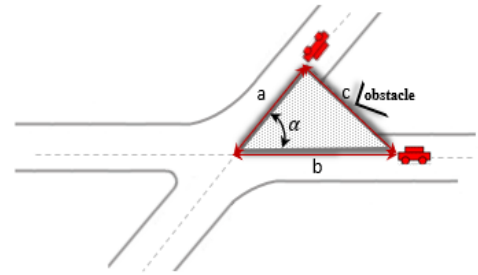

Fig. 5. Illustration of a sight triangle, legs (a, b, c), and angle(α) at non-orthogonal intersection.

$$a \text{ (or, } b) = 1.47 \times V_{posted} \times t_g, \quad (18)$$

where $V_{posted}$ is the road posted speed limit (from 25 to 60 mph in 5 mph), and $t_g \geq 3$ sec is the time gap to enter the other road. The triangle leg (c) is the corner sight distance from a driver eye position on a road to the closest object located on the other road, this distance should be clear out from obstacles. It is also given by AASHTO standards in Table I.

Table I: Lengths of leg (c), and intersection angle (α) at different posted speed limits

| Speed (mph) | 25 | 30 | 35 | 40 | 45 | 50 | 55 | 60 |
|---|---|---|---|---|---|---|---|---|
| c (ft.) | 280 | 355 | 415 | 470 | 530 | 590 | 645 | 705 |
| α ∈ A (deg.) | 60° | 65° | 70° | 75° | 78° | 80° | 85° | 88° |

With the given sight distances (a, b, c), we further explore the possible values of angle α, given in Table I, using a trigonometric function, i.e., $\cos\alpha = \frac{a^2 + b^2 - c^2}{2\,ab}$. Moreover, we investigate the curve of the digamma function within the feasible ranges of the ration of $\left(\frac{D}{h}\right)$ and (1-cos α). The regression analysis shown in Fig. 6 demonstrates that the logarithmic distribution still fits the corresponding digamma functions under different α values. Consequently, substituting the digamma functions by their approximation functions in Eq. (10), we obtain Eqs. (19)–(26) to estimate the upper bound of interference surrogate $\Lambda_i$ occurs at non-orthogonal intersections with different α angles.

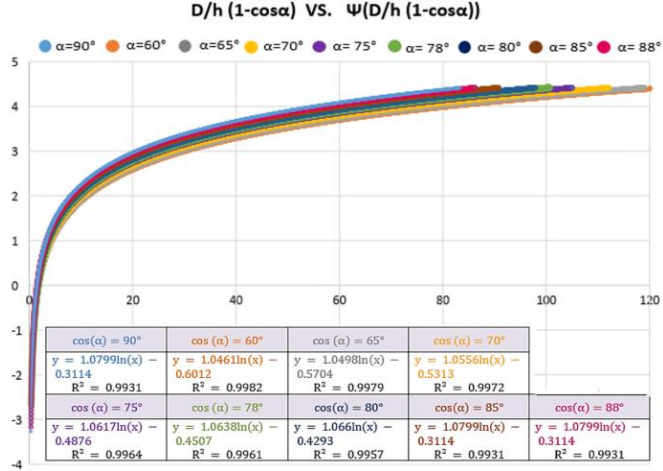

Fig. 6. (D/h (1-cosα)) ratio vs. Digamma function Ψ (D/h (1-cosα)) for a non-orthogonal intersection with angle α

$$\Lambda_i(x) \leq \frac{5}{D^2} + \frac{1}{h^2}[\frac{\pi^2}{6} + 1.3003 \left(\frac{D}{h}\right)^{-1.067}] + [\frac{8}{D^3 h} \times \left(\frac{2h}{D} + 1.0461 \ln\left(\frac{D}{h}(1-\cos\alpha)\right) - 0.024\right)^2 ; \alpha=60° \quad (19)$$

$$\Lambda_i(x) \leq \frac{1}{D^2}[1 + \frac{1}{0.2887}] + \frac{1}{h^2}[\frac{\pi^2}{6} + 1.3003 \left(\frac{D}{h}\right)^{-1.067}] + [\frac{1}{0.1666\, D^3 h} \times \left(\frac{0.5774\, h}{D} + 1.0498 \ln\left(\frac{D}{h}(1-\cos\alpha)\right) + 0.0068\right)^2 ; \alpha=65° \quad (20)$$

$$\Lambda_i(x) \leq \frac{1}{D^2}[1 + \frac{1}{0.3289}] + \frac{1}{h^2}[\frac{\pi^2}{6} + 1.3003 \left(\frac{D}{h}\right)^{-1.067}] + [\frac{1}{0.2165\, D^3 h} \times \left(\frac{0.6579\, h}{D} + 1.0556 \ln\left(\frac{D}{h}(1-\cos\alpha)\right) + 0.0459\right)^2 ; \alpha=70° \quad (21)$$

$$\Lambda_i(x) \leq \frac{1}{D^2}[1 + \frac{1}{0.3705}] + \frac{1}{h^2}[\frac{\pi^2}{6} + 1.3003 \left(\frac{D}{h}\right)^{-1.067}] + [\frac{1}{0.2746\, D^3 h} \times \left(\frac{0.7412\, h}{D} + 1.0617 \ln\left(\frac{D}{h}(1-\cos\alpha)\right) + 0.0896\right)^2 ; \alpha=75° \quad (22)$$

$$\Lambda_i(x) \leq \frac{1}{D^2}[1 + \frac{1}{0.3706}] + \frac{1}{h^2}[\frac{\pi^2}{6} + 1.3003 \left(\frac{D}{h}\right)^{-1.067}] + [\frac{1}{0.2747\, D^3 h} \times \left(\frac{0.7921\, h}{D} + 1.0638 \ln\left(\frac{D}{h}(1-\cos\alpha)\right) + 0.1265\right)^2 ; \alpha=78° \quad (23)$$

$$\Lambda_i(x) \leq \frac{1}{D^2}[1 + \frac{1}{0.4131}] + \frac{1}{h^2}[\frac{\pi^2}{6} + 1.3003 \left(\frac{D}{h}\right)^{-1.067}] + [\frac{1}{0.3414\, D^3 h} \times \left(\frac{0.8263\, h}{D} + 1.066 \ln\left(\frac{D}{h}(1-\cos\alpha)\right) + 0.1479\right)^2 ; \alpha=80° \quad (24)$$

$$\Lambda_i(x) \leq \frac{1}{D^2}[1 + \frac{1}{0.4564}] + \frac{1}{h^2}[\frac{\pi^2}{6} + 1.3003 \left(\frac{D}{h}\right)^{-1.067}] + [\frac{1}{0.4166\, D^3 h} \times \left(\frac{0.9128\, h}{D} + 1.0727 \ln\left(\frac{D}{h}(1-\cos\alpha)\right) + 0.2016\right)^2 ; \alpha=85° \quad (25)$$

$$\Lambda_i(x) \leq \frac{1}{D^2}[1 + \frac{1}{0.4825}] + \frac{1}{h^2}[\frac{\pi^2}{6} + 1.3003 \left(\frac{D}{h}\right)^{-1.067}] + [\frac{1}{0.4657\, D^3 h} \times \left(\frac{0.9651\, h}{D} + 1.0774 \ln\left(\frac{D}{h}(1-\cos\alpha)\right) + 0.2380\right)^2 ; \alpha=88° \quad (26)$$

Again, by applying the MP similar to Eq. (17) but factoring the value of α, we can obtain the tightest upper bounds of $\Lambda_i$ in Eqs. (19)–(26). Eq. (27) illustrates an example for the case α = 60°.

$$\Lambda_{min} = \text{Min } \Lambda_i(x) = \frac{5}{D^2} + \frac{1}{h^2}[\frac{\pi^2}{6} + 1.3003 \left(\frac{D}{h}\right)^{-1.067}] + [\frac{8}{D^3 h} \times \left(\frac{2h}{D} + 1.0461 \ln(\frac{D}{h}(1-\cos\alpha)) - 0.0240\right)^2 \quad (27)$$

$s.t$:
$1.5 < h \leq 86$,
$28 \leq D \leq 125$,
$\alpha = 60°$

These tightest upper bounds for both orthogonal (non-orthogonal) intersections will be used to estimate the most conservative transmission range $r_b$ to sustain a successful V2V transmission through DSRC.

### D. Interference for intersection with multiple lanes:

The above formulations are developed using a traffic intersection with arms having only single lane in each direction. This section next extends our approach to address a traffic intersection with the arms having multi-lanes in each direction. Without loss generality, we demonstrate our approach using the example shown in Fig. 7, in which each arms have two lanes in each direction. First, we label the lanes on each arm by number 1, 2, 3, 4 from the right to left on each arm. And then denote the distance from each lane to the receiver at the neighborhood arm as $L_1$, $L_2$, $L_3$ and $L_4$ respectively. Let distance $L_3$ corresponds to the distance $(x_{ji})$ in the single lane case in Fig. 2. Clearly, the distance $(x_{ji})$ for j vehicles aligned on each other lane can be approximate with respect to the distance $L_3$ as follows: It is known that the ratio of $L_3/L_1$ is equals to the cosine ratio of their horizontal angles ($\frac{\theta_1}{\theta_3}$) as explained in Eq. (28). The similar relationship holds for the second and fourth lane ($L_2$, $L_4$).

$$\Lambda_i^1 = \sum_{j=1}^n 1/\left(\left(\frac{2-\theta_3^2}{2-\theta_1^2}\right) \times x_{ji}\right)^2 = \left(\frac{2-\theta_1^2}{2-\theta_3^2}\right)^2 \sum_{j=1}^n 1/(x_{ji})^2 \quad (28)$$

Next, considering the lane horizontal angle $\theta_l$ become much smaller when more multiple lanes exist, we can use the approximation in [41], $\cos\theta = 1 - \frac{\theta^2}{2!}$ and $\frac{L_3}{L_1} = \frac{2-\theta_1^2}{2-\theta_3^2}$. Next, substituting the approximation function in Eq. (28), and using the interference $\Lambda_i$ given in Eq. (2), we obtain Eq. (29) that represent the interference surrogate $\Lambda_i^1$ from the first lane to a receiver i at the other arm. Similar to Eq. (29), by approximate the distances $L_2$, $L_4$ as to $L_3$, we can obtain the interference $\Lambda_i$ for the second and fourth lane as in Eq. (30), below.

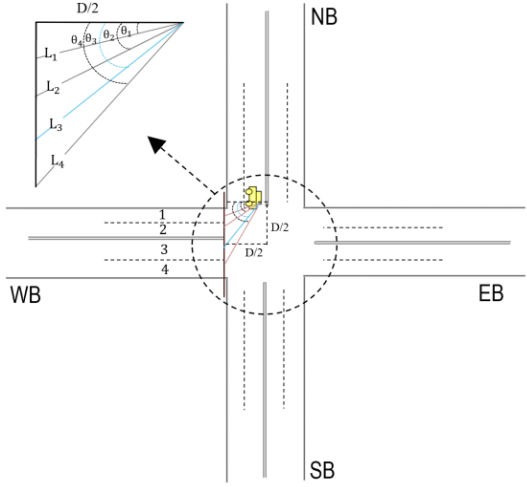

Fig. 7. Inference $\Lambda_i$ at multiple lanes intersection Arm

$$\Lambda_i^1 = \sum_{j=1}^n 1/\left(\left(\frac{2-\theta_3^2}{2-\theta_1^2}\right) \times x_{ji}\right)^2 = \left(\frac{2-\theta_1^2}{2-\theta_3^2}\right)^2 \sum_{j=1}^n 1/(x_{ji})^2 \quad (29)$$

$$\Lambda_i^2 = \left(\frac{2-\theta_2^2}{2-\theta_3^2}\right)^2 \sum_{j=1}^n 1/(x_{ji})^2,$$

$$\Lambda_i^4 = \left(\frac{2-\theta_4^2}{2-\theta_3^2}\right)^2 \sum_{j=1}^n 1/(x_{ji})^2 \quad (30)$$

Furthermore, to obtain the overall interference ($\Lambda_{i_{overall}}$) coming from all lanes of the intersection arm, we use the summation of interference involves the effects from each lane on the intersection arm as given in Eq. (31). Given this summation, we substitute the interference surrogates $\Lambda_i^1$, $\Lambda_i^2$, and $\Lambda_i^4$, by their corresponding approximation functions in Eq. (31), we obtain Eq. (32). The formulation in Eq. (32) is simplified to Eq. (33), to estimate the overall interference of the multi lane arm. Note that $\Lambda_i^3$ in Eq. (31) is substituted by $\Lambda_i$ given in Eq. (2), and parameter (k) is the number of lanes in an intersection arm.

$$\Lambda_{i_{overall}} = \Lambda_i^1 + \Lambda_i^2 + \Lambda_i^3 + \Lambda_i^4 \quad (31)$$

$$\Lambda_{i_{overall}} = \left(\frac{2-\theta_1^2}{2-\theta_3^2}\right)^2 \sum_{j=1}^n 1/(x_{ji})^2 + \left(\frac{2-\theta_2^2}{2-\theta_3^2}\right)^2 \sum_{j=1}^n 1/(x_{ji})^2 + \sum_{j=1}^n 1/(x_{ji})^2 + \left(\frac{2-\theta_4^2}{2-\theta_3^2}\right)^2 \sum_{j=1}^n 1/(x_{ji})^2 \quad (32)$$

$$\Lambda_{i_{overall}} = \sum_{j=1}^n 1/(x_{ji})^2 \left(1 + \sum_{m=1}^k \left(\frac{2-\theta_m^2}{2-\theta_3^2}\right)^2\right) \Rightarrow$$

$$\Lambda_{i_{overall}} = \Lambda_i \left(1 + \sum_{m=1}^k \left(\frac{1-\theta_m^2}{1-\theta_3^2}\right)^2\right) \quad (33)$$

With the above approach to capture the interference at multi lane intersection, the formulation in Eq. (33) is adaptive to the orthogonal and non-orthogonal intersections, by applying different upper bounds of $\Lambda_i$ as in Eq. (16), and Eqs. (19)–(26) for both orthogonal (non-orthogonal) intersections respectively, to estimate the overall interference effect from all lanes on an intersection arm.

### E. Conservative upper bound ($r_b$) for a successful transmission range at traffic intersection

Built upon the interference formulation, the study further explores a conservative upper bound ($r_b$) to identify a transmission range which sustains a successful wireless communication between a receiver (i) and transmitter (w) located at a traffic intersection under various traffic congestion conditions, integrating the interference formulation in Eq. (17). To do that, we add another well-accepted assumption to the SINR that considers white Gaussian background noise with $E(N) \approx 0$ [42][41]. Together with the previous assumptions made for developing the interference surrogate in Eq. (2), the standard SINR formulation in Eq. (1) is transformed into Eq. (34). Plugging the tightest upper bound of interference formulation ($\Lambda_{min}$) obtained from the MP in Eq. (17), Eq. (34) provides us an upper bound ($r_b$) for a successful transmission between the receiver (i) and transmitter (w) both at the stop lines in their respective intersection arm.

$$r \leq r_b = \sqrt[2]{\frac{\beta+1}{\beta} \times \frac{1}{\Lambda_{min}}} \quad (34)$$

**Proposition 2:** The transmission range ($r_b$) in Eq. (34) represents the most conservative upper bound for the transmission range to sustain a successful V2V transmission between any pair of transmitter and receiver at a general traffic intersection under various traffic congestion levels.

Proof: We prove Proposition 2 considering a receiver (i) located at the stop line (see Fig. 1), or at a site further away from the intersection, as shown in Fig. 8 If the receiver (i) is located at the stop line, the tightest upper bound $\Lambda_{min}$ btained from the MP represents the maximum interference at a general traffic intersection with any possible D and α values under various traffic congestion levels (i.e., different values of the spacing h). Accordingly, Eq. (34) which taking the maximum interference leads to the most conservative upper bound for the transmission range.

Next, we demonstrate that a receiver (i′) is exposed to a weaker interference if it is located further away from the stop line of the intersection, such as $x_{i'}(t) = \{x_{i'}(t_1), x_{i'}(t_2), x_{i'}(t_3)\} > 0$ on different arms. According to Eq. (2), the interference at a receiver (i′) is inversely proportional to the distances from the receiver to other vehicles. When a receiver (i′) located away from the stop line such as at a location $(x_{i'}(t_1))$ on the NB, the receiver moves further away from vehicles in groups (a, b, c) on arm WB, EB, or SB, but moves closer to vehicles in the group (d). As a result, the interference generated from all vehicles in the groups (a, b, c), except group (d) becomes weaker (i.e., the squared distance between any vehicle in the groups (a, b, c) is larger than the corresponding distances when the receiver located at the stop line). We consider that the dominant effect will make the interference at the receiver (i′) weaker except a special traffic situation that only arm NB has sufficient traffic, but not all other arms have. The weaker interference leads to a larger efficient transmission range. Therefore, we conclude that Eq.

(34), provides the most conservative transmission range at a traffic intersection.

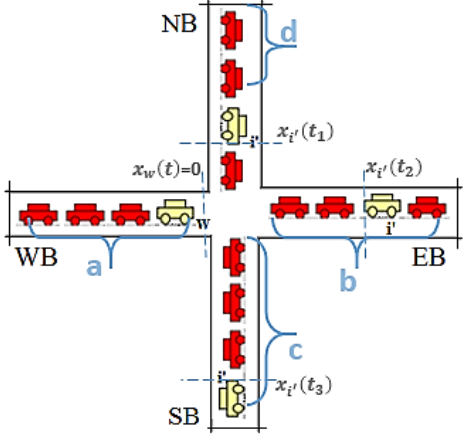

Fig. 8. Receiver ($i'$) located further away from the intersection stop line

Note that the special traffic situation, in which traffic is very sparse at all other arms except the subject arm where a receiver vehicle is located, degrades to a road segment scenario. Existing study in [22] developed an analytical formulation to determine the tightest upper bound for transmission range between a receiver and transmitter vehicles under various traffic flow scenarios.

## IV.  NUMERICAL EXPERIMENTS

This section presents our numerical experiments to validate our theoretical findings and mathematical formulations. The following section presents our testbed, data, experiment design and main results.

### A.  Testbed and Input Data:

Our simulation experiments are built upon two traffic intersections testbeds, which are respectively orthogonal and non-orthogonal intersections (as shown in Fig. 9), in which all approaching arms are two-way with a single lane in each direction. The roadway traffic limits such as average traffic volume, speed limits and geometric features given in Table II are fed in the traffic simulator, VISSIM to generate traffic data in 2400 seconds. The vehicle trajectory data, including the attributes of vehicle ID, average space headway, and average time headway, etc., are collected after the warm-up period (the first 400 seconds) to validate the performance of our approach. The experiments assume all vehicles use same transmission power and broadcast information by flooding.

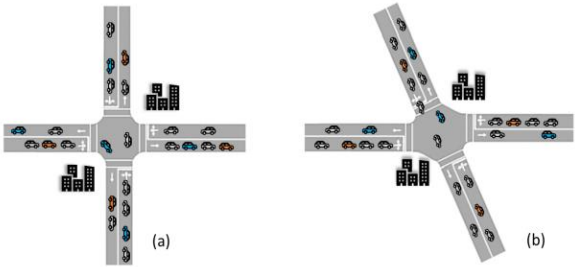

Fig. 9. Two testbed (a) orthogonal intersection (b) non-orthogonal intersection

Table II: General Features of non/orthogonal intersections: entry flows, and geometric parameters

| Orthogonal Intersection | | | | |
|---|---|---|---|---|
| Free flow Speed (mph) | WB-EB 40 | NB-SB 50 | D $\alpha$ | 40 ft. 90° |
| volume (veh/hr/ln) | WB 80 | EB 80 | NB 170 | SB 170 |
| Non-orthogonal intersection | | | | |
| Free flow Speed (mph) | WB-EB 40 | NB-SB 50 | D $\alpha$ | 60 ft. 60° |
| volume (veh/hr/ln) | WB 80 | EB 80 | NB 170 | SB 170 |

### B.  Experiment Design

This study runs multiple experiments, in which the average spacing headway (h) varies from 15 ft to 175 ft to cover different traffic congestion levels, including sparse traffic condition (such as under LOS A or B congestion levels with traffic density <10 veh/$10^3$ft/arm or an average traffic spacing h>100 ft), mild or light traffic congestion (such as under LOS C or D with traffic density 10-20 veh/$10^3$ft/arm or an average traffic spacing h ranges between [50-100)ft]), and dense traffic congestion (such as under LOS E or F with traffic density >25 veh/$10^3$ft/arm, or an average traffic spacing h ranges between (0-50) ft).

The details of the experiments setup and design to evaluate the accuracy of our formulations of interference and the corresponding efficient transmission range are given as follows.

#### a.  Ground Truth

With the collected traffic data, we first evaluate the accuracy of our formulations of interference under different traffic congestion levels, using the results obtained from the SINR as the ground truth. Namely, using the formulation $I$ employed in Eq. (1), the ground truth scenario of our experiments measure the interference according to individual vehicle position at traffic intersection, given the vehicle trajectory data obtained from VISSIM simulation.

Besides the interference obtained from simulation, we measure the interference using our mathematical formulations factoring average traffic spacing and intersection geometric features at traffic intersections. The accuracy is evaluated by the mean absolute percentage error (MAPE denoted as e%) given in Eq. (35), below.

$$e\% = \frac{1}{T}\sum_{t=1}^{T}\frac{|\Lambda_i^M(t)-\Lambda_i^{SINR}(t)|}{\Lambda_i^{SINR}(t)}\times 100\% \qquad (35)$$

where T represents the number of time steps; $\Lambda_i^{SINR}(t)$ represents the interference measured by formulation $I$ used in the SINR in a time step (t), $\Lambda_i^M$ represents the interference measured by the mathematical estimation in Eqs. (6)–(9).

#### b.  The Corresponding Efficient Transmission Range

Given the interference measurement under various traffic conditions, this study next validates the corresponding efficient transmission range estimated by the formulation in Eq. (34) under different traffic congestion levels, including sparse, mild and heavy congestion conditions. More exactly, our experiments measure the upper bound ($r_b$) that identifies a transmission range which sustains a successful wireless communication at traffic intersection, integrating the interference measurements and using β = 0.15 as recommended by [31]. Our experimental results and comprehensive discussions are provided in the section below.

*C. Experiment Results and Discussions*

Built upon the data collected in the numerical experiments, the accuracy of our findings and mathematical formulations is evaluated from two aspects (i) the errors of the interference obtained from the mathematical formulations to the ground truth according to formulation $I$ employed in SINR obtained from the simulation at the orthogonal and non-orthogonal intersections; (ii) the consistency between the corresponding efficient transmission ranges integrating the interference measured mathematically or by simulation along traffic congestion evolvement. The mean absolute percentage error (MAPE) in Eq. (34) is used as the measurement.

Furthermore, this study explores the impact of the critical geometric features (such as (D, α)) at traffic intersections on the V2V communication. Specifically, the experiments test the interference and the corresponding transmission range at traffic intersections with different values of the intersection diameter D and intersection angle (α). The intersection diameter D varies within the range of 30 to 120 ft and the intersection angle (α) varies within 60° to 90° under different congestion levels.

The following sections discuss the experiment results in details.

*a. Interference and Corresponding Transmission Range*

We first demonstrate the performance of the mathematical formulations for approximating the interference at the orthogonal and non-orthogonal intersections respectively. Specifically, the results in Fig. 10, and Fig. 11 indicate that the interference estimated by the mathematical formulations, using macroscopic traffic characteristic (such as the average spacing h), is very close to the formulation $I$ used in the SINR, involving detailed vehicle trajectory data at each time step. More precisely, the mathematical formulations lead to a MAPE equals to 6.2% for the orthogonal intersection and 5.4% for non-orthogonal intersection. Therefore, the mathematical formulations provide accurate estimations to the intersection signal interference using macroscopic traffic characterizes and intersection geometric data rather than detailed vehicle trajectory data. Moreover, we also noticed that when the traffic congestion varies from very sparse (such as under LOS A or B) to mild or light congestion (such as under LOS C or D), the interference at a traffic intersection increases very mildly. However, when the traffic congestion level reaches to LOS E or F, (i.e., heavily congested), the interference ($\Lambda_i$) increases quickly and significantly. These interesting observations will help design efficient MAC algorithms to adapt different traffic conditions.

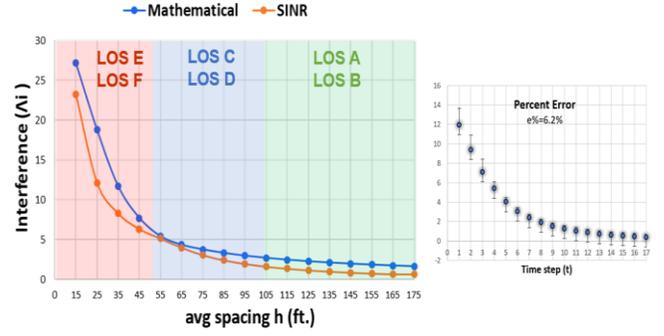

Fig. 10. $\Lambda_i^M$ vs. $\Lambda_i^{SINR}$ at the orthogonal intersection

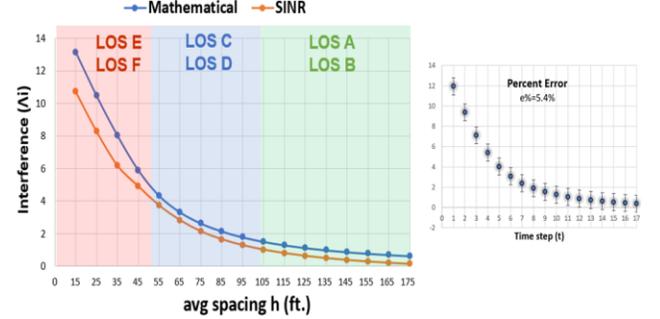

Fig. 11. $\Lambda_i^M$ vs. $\Lambda_i^{SINR}$ at the non-orthogonal intersection

Next, our experiments validate the efficient transmission range ($r_b$) along with traffic congestion evolvement. The results in Fig. 12 and Fig. 13 demonstrate that sparse/dense traffic around intersection leads to a large/small transmission range. This observation is consistent with the formulation in Eq. (34). More importantly, the results indicate that the transmission range ($r_b$) almost linearly increases with the increasing of the average spacing, but the increase rate under LOS E-F is sharper than under LOS A-B or LOS C-D. More exactly, for orthogonal traffic intersection, a sharper slope presents under heavy congestion condition such as LOS E and F than that under sparse and mild traffic such as LOS A-D (see Fig. 12). While for non-orthogonal intersection, a flatter slope presents very heavy congestion condition such as LOS F than under sparse and mild traffic congestion conditions LOS A-D (see Fig. 13). These sophisticated observations demonstrate the significance of factoring traffic conditions and intersection features for better understanding V2V communication features under urban transportation conditions.

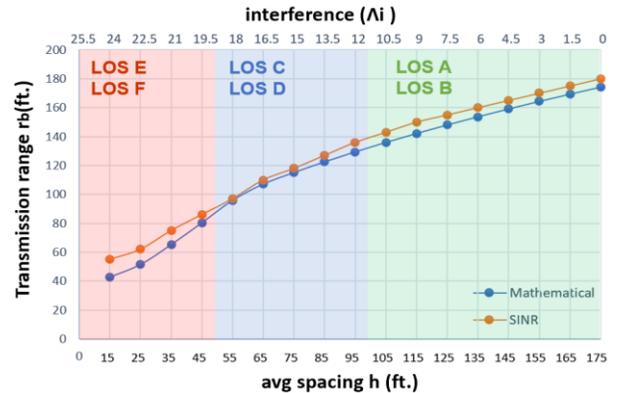

Fig. 12. $r_b$ under different congestion levels at the orthogonal intersection

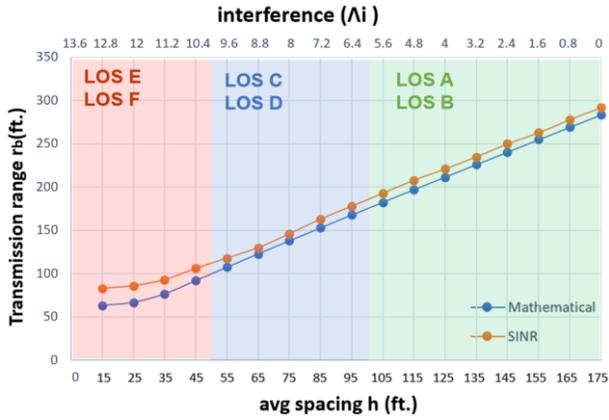

Fig. 13. $r_b$ under different congestion levels at the non-orthogonal intersection

*b. Impact of geometric features on V2V communication*

Furthermore, our experiments examine how the changes in intersection diameter (D) and the intersection angle (α) impacts the interference and the corresponding transmission range at traffic intersections under different traffic congestion levels. The results in Fig. 14 to Fig. 17 demonstrate the impact from intersection diameter (D) on the interference and the transmission range at the orthogonal and non-orthogonal intersections. More exactly, the results in Fig. 14, and Fig. 15 indicate that the interference at a small intersection (i.e., with a small diameter D) is more severe than at a larger intersection (i.e., with a large diameter D), but this difference degrades as the traffic congestion changes from heavily congested to sparse. Fig. 16, and Fig. 17 demonstrate the consistent observations for the transmission range ($r_b$) variation.

On the other hand, Fig. 18 and Fig. 19 show the experiment results regarding the impact of the intersection angle (α) on the interference and the transmission range respectively at the orthogonal and non-orthogonal intersections. Specifically, Fig. 18 demonstrates that the increasing of the intersection angle (α) is prone to increase the interference under all traffic congestion conditions (LOS C-F) but the effect is not apparent under very sparse traffic conditions (LOS A-B). Meanwhile, Fig. 19 demonstrates that a larger intersection angle (α) always leads to a smaller transmission range ($r_b$) but the effect is not significant if the traffic is under very sparse traffic conditions (LOS A-B). One more interesting result shown in Fig. 18 and Fig. 19 is that the orthogonal intersection (i.e. with α=90$^0$) gives the severest interference and minimum transmission range under all different traffic conditions as compared to all other non-orthogonal intersections. Therefore, the case of orthogonal intersection represents a critical case that gives important thresholds for understanding the interference and transmission range as we study the V2V communication performance at an urban traffic intersection.

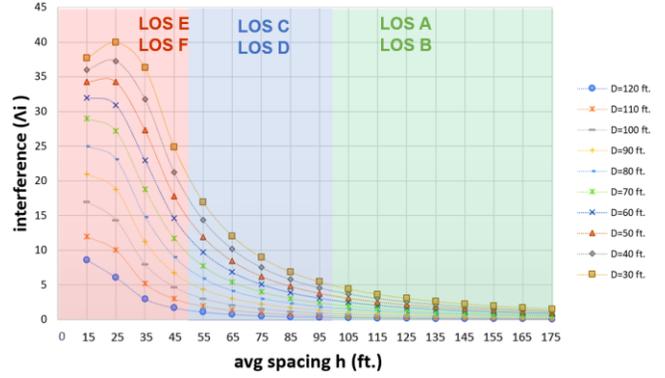

Fig. 14. Interference ($\Lambda_i$) under different D values (orthogonal intersection)

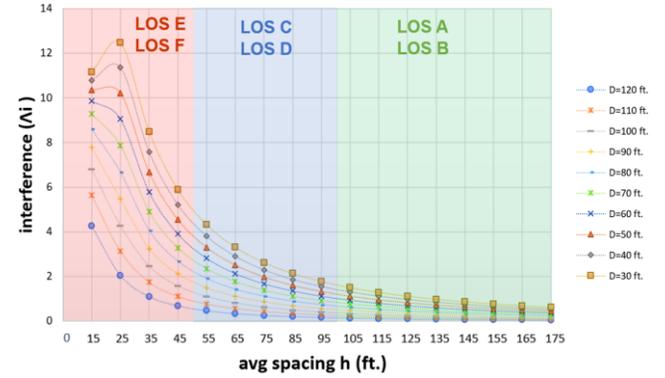

Fig. 15. Interference ($\Lambda_i$) under different D values (non-orthogonal intersection)

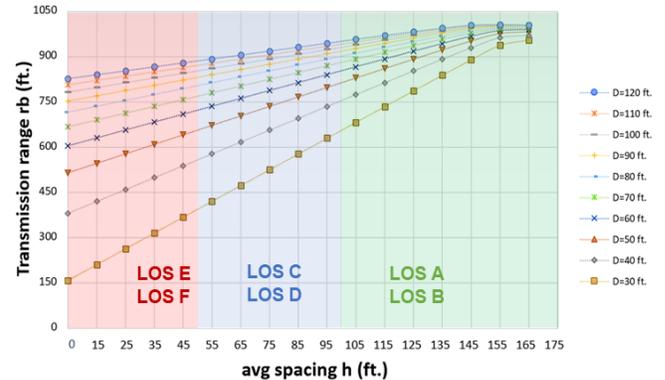

Fig. 16. $r_b$ under different D values (orthogonal intersection)

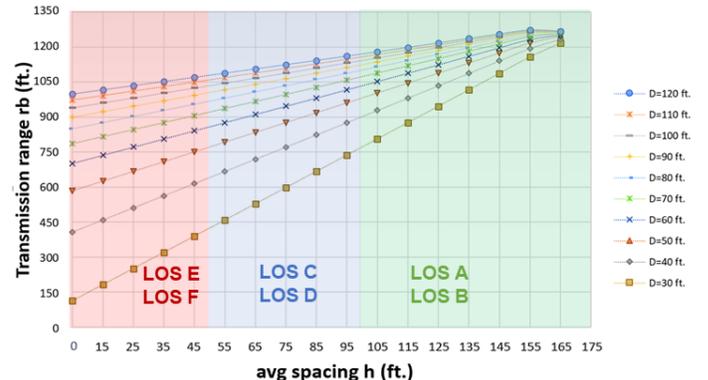

Fig. 17. $r_b$ under different D values (non-orthogonal intersection)

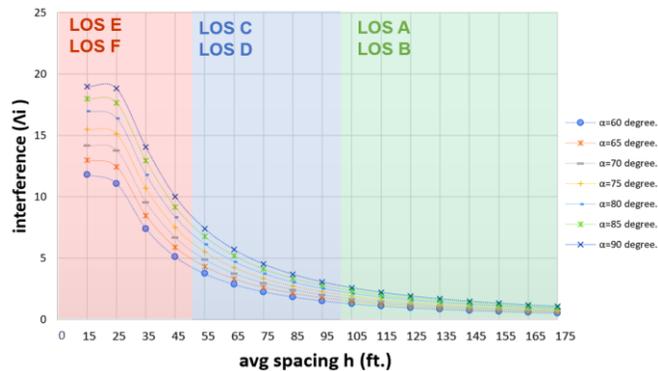

Fig. 18. Interference ($\Lambda_i$) under different angle α values.

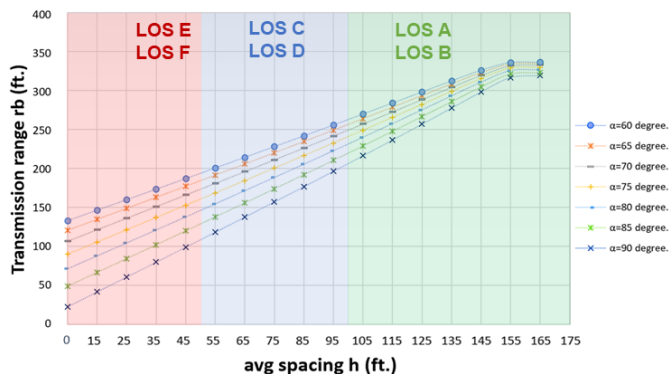

Fig. 19. $r_b$ under different angle α values.

Wrapping up the above observations, we conclude that the mathematical formulations developed in this study work accurately and can help to better understand the signal interference and efficient transmission range of V2V communication occurring at traffic intersection incorporating the impact of traffic congestions and intersection geometric features.

## V. CONCLUSION

This research studies the interference resulting from V2V communication occurring at a traffic intersection, assuming all equipped vehicles broadcast information by flooding. The presented efforts developed mathematical formulations to capture the interference, taking account of macroscopic traffic flow conditions (i.e., average spacing headway $h$) and the critical road geometric features $(D, \alpha)$. Built upon that, we further developed the formulations to approximate a conservative transmission range, which sustains the successful transmission via V2V communication at a traffic intersection. The numerical experiment results indicate that our formulations provide a reliable estimation for the worst-case interference ($\Lambda_i$), with MAPE equals to 6.2%, and 5.4% for orthogonal/non-orthogonal intersections respectively under various traffic congestion levels. Moreover, the results revealed some interesting observations, for example that the interference increases very mildly when the traffic congestion varies from very sparse (such as under LOS A or B) to mild or light congestion (such as under LOS C or D), but increases quickly and significantly when the traffic congestion level reaches to LOS E or F, (i.e., heavily congested). More importantly, the corresponding transmission range almost linearly increases with the increase of the average spacing h, but with different increasing rate under different traffic congestion levels (LOS A to LOS F) either under orthogonal or non-orthogonal intersections. These sophisticated observations clearly raise the awareness of traffic congestion conditions (such as sparse, mild, and heavy congestion level), to deepen the understanding of the average interference and transmission range of V2V in urban traffic intersections.

Furthermore, our experiments noticed a clear impact from intersection geometric features $(D, \alpha)$ on the V2V communication features under different traffic congestion levels. Specifically, a smaller intersection (i.e., with a smaller diameter D) will result in more severe interference and smaller efficient transmission range under heavy traffic congestion level. Also, a larger intersection angle is prone to result in a more severe interference (shorter transmission range) under mild or heavy traffic congestion conditions, but these effects are not apparent for the interference and the efficient transmission range under sparse traffic conditions (i.e., LOS A or B). one more important observation, that the orthogonal intersection (i.e. with $\alpha = 90^0$) leads to the most severe interference and minimum transmission range under various traffic conditions as compared to all other non-orthogonal intersections. This will potentially help us understand the thresholds of V2V communication performance at other general intersections.

Wrapping up all results, we claim that traffic conditions and intersection features significantly affect the V2V communication under urban transportation conditions. The findings of this study will potentially help in developing efficient MAC algorithms adaptive to urban traffic conditions as well as provide support for the successful implementations of various V2V applications in ITS systems.

## VI. ACKNOWLEDGMENT

This research is partially supported by the National Science Foundation award: #1436786.

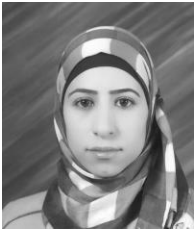**Ala Alobeidyeen** received the B.S. degree in civil engineering from Tafila Technical University, Tafila, Jordan, in 2012 and the M.S. degree in Transportation engineering from University of Jordan, Amman, Jordan, USA, in 2015, she currently working toward the Ph.D. in Transportation engineering in the Department of Civil and Coastal Engineering, University of Florida, Gainesville, FL, USA. Her research interests include vehicle-to-vehicle communication over traffic networks, Transit system and Connected and Autonomous vehicles. She is also a member of the IEEE and Transportation Research Board.

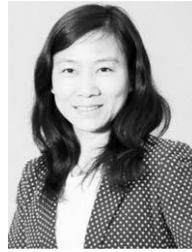**Lili Du** received the B.S. degree in mechanical engineering from Xi'an Jiaotong University, Xi'an, China, in 1998; the M.S. degree in industrial engineering from Tsinghua University, Beijing, China, in 2003; and the Ph.D. degree from Rensselaer Polytechnic Institute, Troy, NY, USA, in 2008. She is currently an Associate Professor with the Department of Civil and Coastal Engineering, University of Florida, Gainesville, FL, USA. Her work has appeared in the Association for Computing Machinery Workshop on Vehicular Ad Hoc Networks publication, Transportation Research Part B: Methodological, Transportation Research Part C: Emerging Technologies, Networks and Spatial Economics, Renewable Energy, and International Journal of Production Research, among others. Her research interests include vehicle-to-vehicle communication networks, connected vehicles, intelligent transportation systems, real-time traffic sensing, and transportation network design and system analysis. Dr. Du is a member of the IEEE and Transportation Research Board Transportation Network Modeling Committee (ADB30).